\documentclass[preprint]{aastex}

\newcommand{\msun}{M$_\odot$}

\newcommand{\days}{$d$}
\newcommand{\degree}{$^\circ$}

\slugcomment{}
\shorttitle{Mid-IR Observations of the Mira Circumstellar Environment}
\shortauthors{M. Marengo et al.}

\begin{document}


\title{Mid-IR Observations of Mira Circumstellar Environment}

\author{Massimo Marengo, Margarita Karovska, Giovanni G. Fazio, Joseph
L. Hora}
\affil{Harvard-Smithsonian Center for Astrophysics, 60 Garden St.,
Cambridge, MA 02138}
\email{mmarengo@cfa.harvard.edu}

\author{William F. Hoffmann}
\affil{Steward Observatory, University of Arizona, 933 North Cherry
Avenue, Tucson, AZ 85721-0065}

\author{Aditya Dayal}
\affil{KLA-Tencor Corporation, 160 Rio Nobles, San Jose, CA 94134}

\and
\author{Lynne K. Deutsch}
\affil{Department of Astronomy, CAS 519, Boston University, 725
Commonwealth Ave., Boston, MA 02215}


\begin{abstract}
This paper presents results from
high-angular resolution mid-IR imaging
of the Mira AB circumbinary environment using the MIRAC3 camera
at the NASA Infrared Telescope
Facility (IRTF). 
We resolved the dusty circumstellar envelope
at 9.8, 11.7 and 18~\micron{} around Mira A (o Ceti),  and measured
the size of the extended emission.  
Strong deviations from spherical symmetry are detected in the images
of Mira AB system, 
including possible dust clumps in the direction of the companion (Mira
B).
These observations
suggest that Mira B plays an active role in shaping
the morphology of the circumstellar environment of
Mira A as it evolves toward the Planetary Nebula phase.
\end{abstract}




\section{Introduction}\label{sec-intro}

\objectname[]{Mira A} ($o$ Ceti, Mira) is a cool pulsating giant of M2-7 III
spectral type, with a mass comparable to our Sun and a diameter of
several hundred solar radii. Prototype of the Mira-type class of Long
Period Variables (LPVs), it has a period of 332\days{} during which
its brightness changes by several magnitudes in the optical. 
Mira has a hot companion (Mira B) at an angular distance of $\sim
0.6$''. The companion is probably a white dwarf accreting from Mira's
wind \citep{reimers1985, karovska1997}.

Recent HST observations detected significant asymmetries in the
giant's atmosphere and found evidence for possible interaction with
its companion. 
The HST Faint Object Camera (FOC) optical 
images showed that Mira A atmosphere
is elongated in the direction of 175\degree{} with an apparent
size of  56~mas. HST UV images showed a ``hook-like'' structure
extending eastward from Mira A photosphere towards Mira B \citep{nasa1997, 
karovska1997}.
These results suggest that the companion may have an important role in
shaping the circumstellar environment of the system. 

Mira A is losing mass via a strong dust driven wind, at the rate of
$\sim 5 \cdot 10^{-7}$ \msun~yr$^{-1}$ \citep[see][and references
therein]{loup1993}. As a consequence, the star is surrounded by an
extended circumstellar envelope of gas and dust. 
Observations of Mira A molecular envelope at radio wavelengths
show deviations from
spherical symmetry \citep[see e.g.][]{planesas1990,josselin2000}.
The dust is revealed
via its strong infrared excess \citep{iras1986, gezari1993}, dominated
by an emission feature at 10~\micron{} which identify an O-rich
composition \citep{monnier1998}. 
Evidence for spatially extended emission from Mira circumstellar dust
was first obtained by \citet{bauer1994}, analyzing the IRAS satellite
data at 60~\micron. More recently, mid-IR observations have been 
performed at 11~\micron{} with the ISI interferometer, 
probing the dust distribution at a smaller spatial scale. By fitting the
observed visibilities with models, \citet{lopez1997} inferred
the presence of asymmetries and inhomogeneities in Mira's dust
envelope. 

Mira-type stars are in the Asymptotic Giant Branch (AGB) phase, and
are among the precursors of Planetary Nebul\ae{} \citep[see][for a
review]{habing1990}. 
The dynamical evolution of AGB circumstellar envelopes is largely
controlled by the dust component, which mediates the momentum transfer
from the radiation field to the molecular gas. Mapping the spatial
distribution of the dust grains in the Mira AB circumbinary environment
is thus necessary to understand the future evolution of the system 
towards the PN phase, and the dynamics of the interaction between its
two stellar components. 

We present here the first
sub-arcsecond mid-IR images of Mira A circumstellar environment
which map the 2-dimensional
distribution of the dust emission. The
observations and the techniques for data acquisition and reduction are
presented in the next section. The results are shown in
section~\ref{sec-discuss}, and discussed in relation to other
available observations and models.


\section{Observations and Data Reductions}\label{sec-obs}

The observations were performed on September 22, 1999 (JD2451444) when
the source was at light curve phase 0.7, close to minimum luminosity
(AAVSO). We used the mid-IR camera MIRAC3
\citep{hoffmann1998}, mounted at the 3.0m NASA Infrared Telescope
Facility (IRTF). MIRAC3 uses a Boeing HF16 128$\times$128 Si:As
blocked impurity band detector. On the IRTF MIRAC3 has a plate scale
of 0.34 arcsec pix$^{-1}$, providing a total field of view of
42''$\times$42''. This pixel scale ensures Nyquist sampling of the
diffraction-limited Point Spread Function (PSF). 

We obtained images of Mira AB in MIRAC3 9.8, 11.7 and 18.0~\micron{} 10\%
passband filters, with a total on-source integration time of 1600~s at
each wavelength. The reference star $\alpha$~Tau was observed when
transiting at a similar airmass of the source, to provide flux and PSF
calibration.

We used a standard nodding and chopping technique to remove the
background signal, dithering the source on the array to obtain 
sub-sampling of the PSF. The chop frequency was set to 2~Hz, with a
throw of 20'' in the north-south direction. The nod throw was also set
to 20'', but in the east-west direction, in order to have all four
chop-nod beams inside the field of view of the array. Each individual
nod cycle required 10~s on-source integration, and the procedure was
repeated for as many cycles as needed to obtain the requested total
integration time.

The data was analyzed by first subtracting the chop-on from the
chop-off frames for both nodding beams. The two images thus obtained
were then subtracted one from the other, in order to get a single
frame in which the source appears in all four beams (two negative and
two positive). We then applied a gain matrix, derived from images of
the dome (high intensity uniform background) and the sky (low
intensity uniform background), to flat field the chop-nodded image. 

This procedure was repeated for each of the nodding cycles for which
the source was observed. A final high signal-to-noise ratio cumulative
image was then obtained by co-adding together all the beams, each
re-centered and shifted on the source centroid. The last bit of
co-adding was performed on a sub-pixel grid having the size of
one-fifth (9.8 and 11.7~\micron{} images) or one-third (18.0~\micron{}
image) of the original MIRAC3 pixels, thus providing a final pixel
scale of 0.068 and 0.114 arcsec pix$^{-1}$ respectively. A mask file to
block out the effects of bad pixels and field vignetting was also
created and applied at this stage, preventing individual rejected
pixels from contributing to the final image. The same observing and
reduction procedure was also used for the reference star, to ensure a
uniform treatment of the source and the standard.


\section{Results and Discussion}\label{sec-discuss}

\subsection{Source photometry}

We first estimated the brightness of Mira AB using the multi-wavelength
co-added images. The aperture photometry at the three observed
wavelengths yields $\sim 2710$~Jy at 9.8~\micron, $\sim 2170$~Jy at
11.7~\micron{} and $\sim 2020$~Jy at 18.0~\micron{} (see
Table~1). Photometric errors are of the order of 5\% at 9.8 and
11.7~\micron{} and 8\% at 18.0~\micron. 

Our measurements are consistent with the published mid-IR fluxes of the
source \citep{gezari1993} which are in the range 1300--5400~Jy at
$\sim 10$~\micron, 1500--5500~Jy at 11--12~\micron{} and 1100--2800~Jy
around 18~\micron. This wide spread, in part due to the large
uncertainties of the published data, is a consequence of the source
long period variability, which is still present at infrared
wavelengths. Note that, in the mid-IR, only a small fraction of the
overall luminosity is contributed by Mira A stellar flux, due to the
greater brightness of the dust envelope. The observed infrared
variability is thus affected by the changes in the physical and
thermodynamical status of the circumstellar dust.

\subsection{Size of Mira's dust envelope}

Our images of Mira AB at 9.8, 11.7 and 18.0~\micron{} indicate that the
dusty envelope is spatially resolved at all observed wavelengths. This
is shown in Figure~1, which plots the azimuthal averages of the source
and reference (radial profiles), normalized at the same peak value. In all
panels the source profile has a larger FWHM than the reference star
(assumed to be a point source), which is a clear indication of spatial
extension. The measured FWHM are listed in Table~1.

We estimated the apparent size of Mira AB envelope as a function of
wavelength by convolving the reference star profile with a gaussian
model of Mira A envelope, and then fitting to the source radial profile.
We obtained a FWHM of 0.34'',
0.35'' and 0.86'' for the best fit gaussian model at 9.8, 11.7 and
18.0~\micron{} respectively. 
As expected, the apparent sizes of Mira are almost the same
at 9.8 and 11.7~\micron{}. Mira A envelope is significantly larger at
18~\micron{}. We obtained very similar results using deconvolved
images described in the next section.
Using a Hipparcos measured distance of 128~pc \citep{perryman1997}
we estimate the apparent spatial scale of
Mira A dust envelope from 50~AU (at 10~\micron) to 100~AU at 18~\micron. 

The measured apparent sizes of the dust envelope
are 10-20 times larger then the apparent size of
Mira A photosphere. For example,
the mid-infrared observations using the ISI interferometer resolved
Mira A stellar photosphere, measuring a uniform disk diameter of
47.8~mas \citep{weiner2000}. 

The larger size of the 18~\micron{} envelope can be explained on
the bases of the radiative transfer properties of AGB envelopes, which
in  general have a radial thermal structure that can be approximated
with a power-law $T \propto r^{-0.4}$ \citep{ivezic1997}. Since the
peak of emission for the dust radiation at a certain wavelength is
inversely proportional to the grain temperature (Wien's law), one
should expect $T_{18\mu \rm m} / T_{11.7\mu \rm m} \simeq 0.65$. This
means that the size of the region where most of the 18~\micron{} flux
is produced should be $r_{18 \mu \rm m} \simeq 2.9 \, r_{10 \mu \rm
m}$. This factor is similar to the measured ratio between the FWHM of
the 18.0 and 11.7~\micron{} deconvolved images ($\sim 2.5$), and much
larger than the PSF spread due to the telescope resolving power
(proportional to $\lambda$, e.g. $\sim 1.5$). Note that this
calculation is only  a rough estimate for the real size of the
envelope.
Better estimates can be obtained using models including physical
parameters such as the dust condensation radius. This requires a 
simultaneous fitting of the radial intensity profile and
the source spectral energy distribution, which allows derivation of the
envelope optical depth $\tau_V$ and the inner shell temperature $T_1$
\citep{marengo2001}. Results using these models will be presented in a
separate paper.

\subsection{Envelope asymmetries}

Our multi-wavelength direct mid-IR images show asymmetries that aren't
seen in the images of the reference star taken with the same filters. 
We confirmed the presence of significant departures from
spherical symmetry in Mira's dust envelope by performing
deconvolution using the Richardson-Lucy method \citep{richardson1972,
lucy1974}. The images at 9.8 and 11.7~\micron{}
were deconvolved with the corresponding reference star images used as
PSFs. This technique was not applied to the 18.0~\micron{} image, due
to the lower S/N of the source/reference pair.

The deconvolved images are similar at 9.8 and 11.7~\micron, as shown in
Figure~2. 
They show that the spatial distribution of the dust in Mira A
circumstellar environment is clearly not spherical. Two dominant
asymmetries are detected in the images: one with a major axis position
angle of 175\degree, and another at a position angle of
approximately 110\degree.

The first asymmetry is in the same direction as the asymmetry in the
Mira A TiO envelope observed with the HST in 1995
\citep{karovska1997}, but its apparent size is roughly ten times
larger then the one measured in the HST images.
Near-IR images obtained in the JHK
bands by \citet{cruzalebes1998} also show a N-S elongated shape, which has
been interpreted as scattered light from dust.
The observed asymmetries could be due to an asymmetric outflow, as
suggested by \citet{planesas1990} and \citet{josselin2000}.

The second dominant asymmetry in Mira AB mid-IR images 
is in the direction of the companion. This is in agreement with
earlier mid-IR observations suggesting the presence of an eastward
asymmetry \citep{lopez1997, meixner2000}. 
Figure~2 shows the location of the companion as observed by the HST
in the late 1995 \citep{karovska1997}. We do not expect that the
position angle and the separation of the companion have changed
substantially since 1995, because of its long orbital period
of several hundred years
\citep{baize1980}. Our images indicate that the companion is
embedded in the elongated dust envelope of Mira A.
Because of the limited angular resolution in these observations
we cannot determine if the asymmetry is due to a second unresolved clump 
located in the vicinity of the companion, or is a continuous
extension of the Mira A dust envelope toward the companion.

Furthermore, a fainter clump slightly above the noise level 
is resolved at a distance of
1.1'' from the main source, with a position angle 115\degree. Its
brightness is $\sim 14.0 \pm 1.5$~Jy at 9.8~\micron{} and $\sim 11.0
\pm 1.0$~Jy at 11.7~\micron, which is a factor $\sim 200$ fainter than
the main source. We estimate a color temperature  of $390 \pm 100$~K,
obtained by black body fitting of the clump brightness at the two
observed wavelengths. 
The presence of dust clumps has been suggested by 
interferometric observations using the ISI 11~\micron{} visibilities
by \citet{lopez1997}. Their ``Clump~3''
model, in particular, invokes the presence of dust clumps with
position angle 120\degree, which may thus be associated with our
observed structures.

The mid-IR observations of Mira AB system indicate that the shape of
Mira A circumstellar envelope may be directly influenced
by the presence of the companion orbiting the main AGB mass losing
star.
Hydrodynamic models in dusty winds of close binary systems
\citep{mastrodemos1999} predict the formation of spiral shocks in the
orbital plane, which trail and follow the secondary star.
The structures observed in the mid-IR may be related to this kind of
phenomenon. 
However, the interpretation of the observed geometry is still open,
given that the dynamics of the interactions between the two components
of close binary systems is not well understood.

\section{Conclusions}\label{sec-conclus}

Our mid-IR images of Mira AB dust envelope 
show significant departures from spherical
symmetry. They also indicate that
the overall distribution of the circumstellar dust
is affected by the presence of a companion. Given the role of dust
in shaping the future dynamical evolution of the system, these
observations not only give a better characterization of the geometry
of the Mira AB binary system, but may also contribute to a better
understanding of the formation of asymmetric PN.

Many details of the transition from AGB to PNs cannot be
explained on the basis of the spherically symmetric ``interacting
stellar winds'' theory \citep{kwok1978} which defines the general
framework of PN formation. High resolution HST images of PNs reveal a
general absence of spherical symmetry, in favor of bipolar structures
enriched by a multitude of micro-structures (jets, clumps, ``fliers'',
etc\dots). 
Optical and mid-IR images of post-AGB and
pre-PN \citep{sahai1998, meixner1999, ueta2000} show envelope
asymmetries already present before the onset of the PN phase. 
How these structures evolve is unknown. The results of our IR
observations show that Mira AB offers the possibility of observing the
beginning of this process.


\acknowledgements
We thank the IRTF staff for their outstanding support. M.~K. is a
member of the Chandra Science Center, which is operated under contract
NAS8-39073, and is partially supported by NASA. We also thanks the
referee for his useful comments.

\clearpage


\clearpage


\begin{table}
\begin{scriptsize}
\begin{center}
\begin{tabular}{lccc}
\multicolumn{4}{c}{\scriptsize TABLE 1}\\
\multicolumn{4}{c}{\scriptsize MID-IR PHOTOMETRY AND ANGULAR SIZE}\\
\hline
\hline
Wavelength & 
Photometry & 
\multicolumn{2}{c}{FWHM radial average profile}\\
& & source & reference\\
\hline
\hline
 9.8~\micron & 2710 ($\pm$ 135) Jy & 1.05'' & 0.97''\\
11.7~\micron & 2170 ($\pm$ 110) Jy & 1.08'' & 1.01''\\
18.0~\micron & 2020 ($\pm$ 270) Jy & 1.56'' & 1.32''\\
\hline
\end{tabular}
\end{center}
\end{scriptsize}
\end{table}

\clearpage


\begin{center}
\includegraphics[width=0.9\textwidth]{fig1.ps}
\end{center}

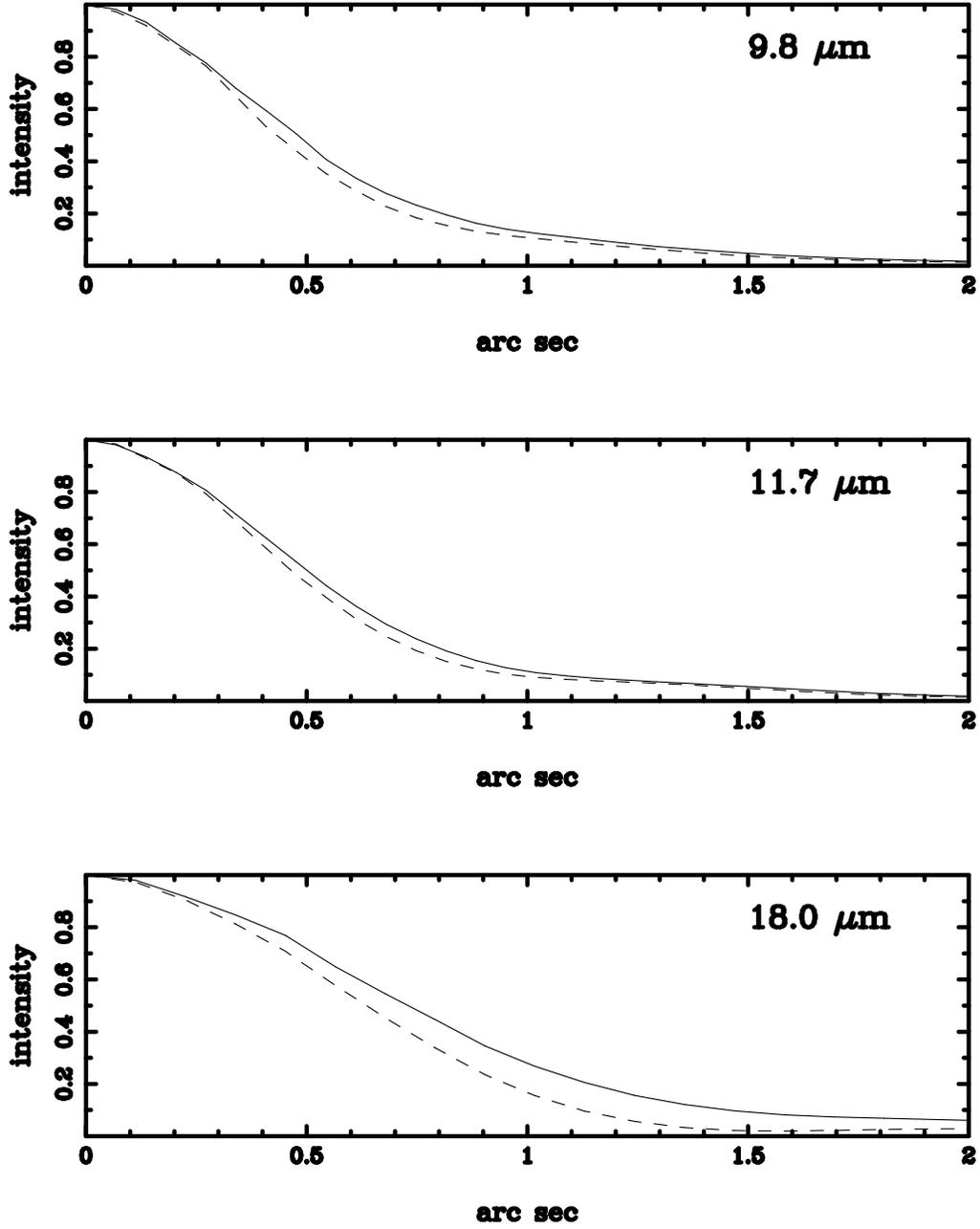
Figure 1. Radial profiles of Mira AB image (solid line) and of the
reference image (dashed line) at 9.8, 11.7 and 18.0~\micron{}.

\clearpage

\begin{center}
\includegraphics[height=0.99\textwidth,angle=-90]{fig2.ps}
\end{center}

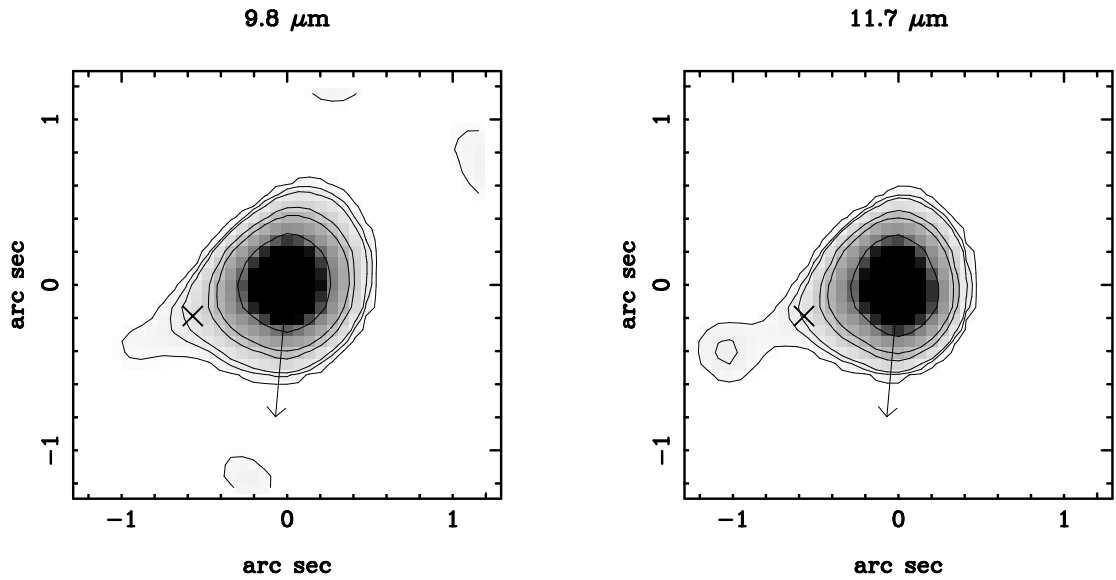
Figure 2. Deconvolved images of Mira AB at 9.8 and 11.7~\micron. North
is up and West is right. The location of Mira B, as observed by the
HST in 1995, is marked by a cross; the arrow indicates the orientation
of the dominant asymmetry in the 1995 HST image of Mira A
\citep{karovska1997}. The contours are plotted for 0.001, 0.005, 0.01,
0.05, 0.1 and 0.5 of the maximum flux density level.

\end{document}